\documentclass[superscriptaddress,twocolumn,prl]{revtex4}

\usepackage{graphicx}
\usepackage{dcolumn}
\usepackage{multirow}
\usepackage{bm}
\usepackage{times}

\usepackage{color}

\begin{document}

\title{Signatures of Electronic Correlations in Optical Properties of LaFeAsO$_{1-x}$F$_x$}

\author{A.V. Boris}
\email[]{A.Boris@fkf.mpg.de}
\affiliation{Max-Planck-Institut f\"{u}r Festk\"{o}rperforschung,
Heisenbergstrasse 1, D-70569 Stuttgart, Germany}
\affiliation{Department of Physics, Loughborough University, Loughborough,
LE11 3TU, United Kingdom}
\author{N.N. Kovaleva}
\affiliation{Max-Planck-Institut f\"{u}r Festk\"{o}rperforschung,
Heisenbergstrasse 1, D-70569 Stuttgart, Germany}
\affiliation{Department of Physics, Loughborough University, Loughborough,
LE11 3TU, United Kingdom}
\author{S.S.A. Seo}
\affiliation{Max-Planck-Institut f\"{u}r Festk\"{o}rperforschung, Heisenbergstrasse
1, D-70569 Stuttgart, Germany}
\author{J.S. Kim}
\affiliation{Max-Planck-Institut f\"{u}r Festk\"{o}rperforschung, Heisenbergstrasse 1, D-70569 Stuttgart, Germany}
\author{P. Popovich}
\affiliation{Max-Planck-Institut f\"{u}r Festk\"{o}rperforschung, Heisenbergstrasse 1, D-70569 Stuttgart, Germany}
\author{Y. Matiks}
\affiliation{Max-Planck-Institut f\"{u}r Festk\"{o}rperforschung, Heisenbergstrasse
1, D-70569 Stuttgart, Germany}
\author{R.K. Kremer}
\affiliation{Max-Planck-Institut f\"{u}r Festk\"{o}rperforschung, Heisenbergstrasse 1, D-70569 Stuttgart, Germany}
\author{B. Keimer}
\affiliation{Max-Planck-Institut f\"{u}r Festk\"{o}rperforschung, Heisenbergstrasse 1, D-70569 Stuttgart, Germany}
\date{\today}
\begin{abstract} 
Spectroscopic ellipsometry is used to determine the dielectric function of the superconducting LaFeAsO$_{0.9}$F$_{0.1}$ ($T_c$ = 27 K) and undoped LaFeAsO polycrystalline samples in the wide range 0.01-6.5 eV at temperatures 10
$\leq T \leq$ 350 K. The free charge carrier response in both samples is heavily damped with the effective carrier density as low as 0.040$\pm$0.005 electrons per unit cell. The spectral weight transfer in the undoped LaFeAsO associated with opening of the pseudogap at about 0.65 eV is restricted at energies below 2 eV. The spectra of superconducting LaFeAsO$_{0.9}$F$_{0.1}$ reveal a significant transfer of the spectral weight to a broad optical band above 4 eV with increasing temperature. Our data may imply that the electronic states near the Fermi surface are strongly renormalized due to electron-phonon and/or electron-electron interactions.

\end{abstract} 
\pacs{74.70.-b, 74.25.Gz, 71.27.+a, 72.80.Ga} 
\maketitle
The iron-based layered oxypnictides LaFeAsO$_{1-x}$F$_x$ represent a new class of superconductors with the highest transition temperature known apart from the cuprates \cite{Kamihara,Takahashi,Cruz,Hunte,Chen}.
The relevance of electronic correlations for the unusual properties of these materials in the normal and superconducting states is being intensely debated \cite{kotliar1,kotliar2,singh,craco,dolgov,mazin,kuroki,lee,yin}, and experiments that directly address this issue are highly desirable.
Recent theoretical and experimental advances have demonstrated that electronic correlations profoundly influence the optical response at energies up to several eV. \cite{kotliar1,kotliar3,toschi,millis,boris,kovaleva}.
Spectroscopic ellipsometry allows one to accurately detect such modifications and is hence a highly distinctive probe of electronic correlations in transition metal oxides \cite{boris,kovaleva}. 
Dynamical mean-field theory (DMFT) calculations explain many aspects of 
strong temperature and doping dependent modifications of the optical spectral
weight (SW) in the normal state of the superconducting cuprates \cite{kotliar3,toschi,millis}.
Very recent DMFT studies indicate that LaFeAsO is slightly below the critical value of the Hubbard $U$ required to obtain an insulating state \cite{kotliar1,kotliar2}. The electronic correlations are predicted to be strong enough to renormalize the density of states of the conduction band in such a way that most of its
SW is transferred to the Hubbard band.

We report a comprehensive spectroscopic ellipsometry study of superconducting LaFeAsO$_{0.9}$F$_{0.1}$ ($T_c = 27$ K) and nonsuperconducing LaFeAsO over a wide range of temperatures (10 - 350 K) and photon energies extending from the far infrared (IR) into the deep ultraviolet (UV), 0.01 - 6.5 eV. The optical conductivity spectra are dominated by a sequence of interband transitions (Fe$3d-$Fe$3d$, As$4p-$Fe$3d$, O$2p-$Fe$3d$, and O$2p-$La$5d$), and the contribution from the free charge carriers to the optical conductivity is extremely weak ($0.04 \pm 0.005$ electrons per unit cell), confirming the DMFT predictions.
Further support for strong electronic correlations is derived from the temperature and doping dependence of the high-energy optical response. Specifically, the spectra of superconducting LaFeAsO$_{0.9}$F$_{0.1}$ reveal a significant transfer of SW from low energies, $\hbar\omega\lesssim $ 2 eV,  to a broad optical band above 4 eV with increasing temperature, whereas the SW redistribution in nonsuperconducting LaOFeAs is restricted to energies below 2 eV. Despite the small ($\sim 10 \%$) difference in doping level, both materials exhibit a significantly different optical response, which is indicative of a profound and unexpected difference in their electronic structure.

\begin{figure}[ht]
\includegraphics[width=7.7cm]{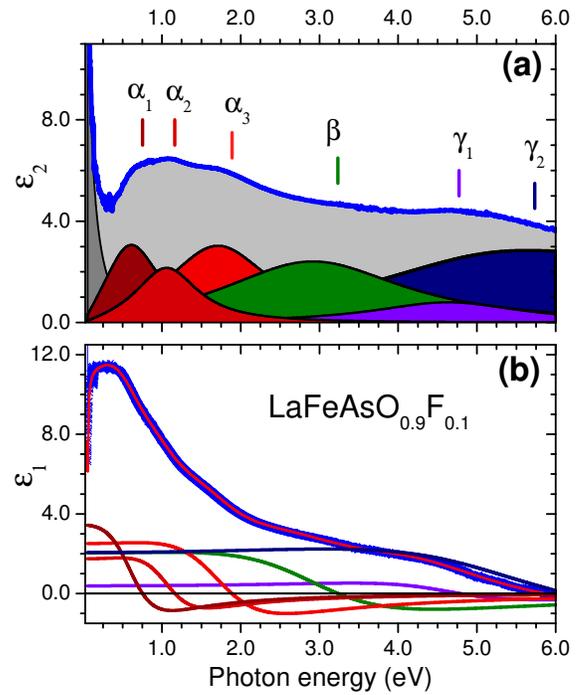} 
\caption{(a) Imaginary and (b) real parts of the dielectric function
of LaFeAsO$_{0.9}$F$_{0.1}$ measured at $T=300$ K (blue heavy lines) and represented by the total contribution [(a) shaded area,(b) red line] of separate Lorenzian bands determined by the dispersion analysis, as described in the text. The dark gray shaded area shows the free charge carrier contribution to $\varepsilon_2(\omega)$.}
\protect\label{fig1} 
\end{figure} 

Polycrystalline samples of pristine and 10$\%$ F-substituted LaFeAsO were prepared following the procedure described in Ref. \onlinecite{Chen}. Their chemical composition was confirmed by energy-dispersive x-ray spectroscopy. Powder x-ray diffraction measurements confirmed that the structure is  tetragonal of ZrCuSiAs-type, with lattice parameters  consistent with previous reports
\cite{Kamihara,Cruz,Chen}. The amount of the impurity phase is estimated to be less then 5$\%$. The magnetic susceptibility and resistivity data show critical behavior at the phase transition temperatures: for the LaFeAsO sample we determine a magnetic ordering temperature $T_{SDW}\approx$ 155  K, whereas the LaFeAsO$_{0.9}$F$_{0.1}$ sample exhibits the onset of superconducting transition temperature at $T_c$ $\approx$ 27 K,  with a transition width of $\sim 3$ K. The bulk-sensitive probes such as magnetic susceptibility, resistivity and IR reflectivity measurements indicate temporary and atmospheric stability of the samples. However, we notice that the visible-UV conductivity of the F-doped sample changes in time at ambient air conditions (likely reflecting out-diffusion of fluorine), and in weeks it approaches the behavior of the undoped LaFeAsO sample. We report our data measured on freshly prepared superconducting LaFeAsO$_{0.9}$F$_{0.1}$. The samples were polished parallel to the broad face just prior to the measurements.

\begin{figure}[ht]
\includegraphics[width=7.7cm]{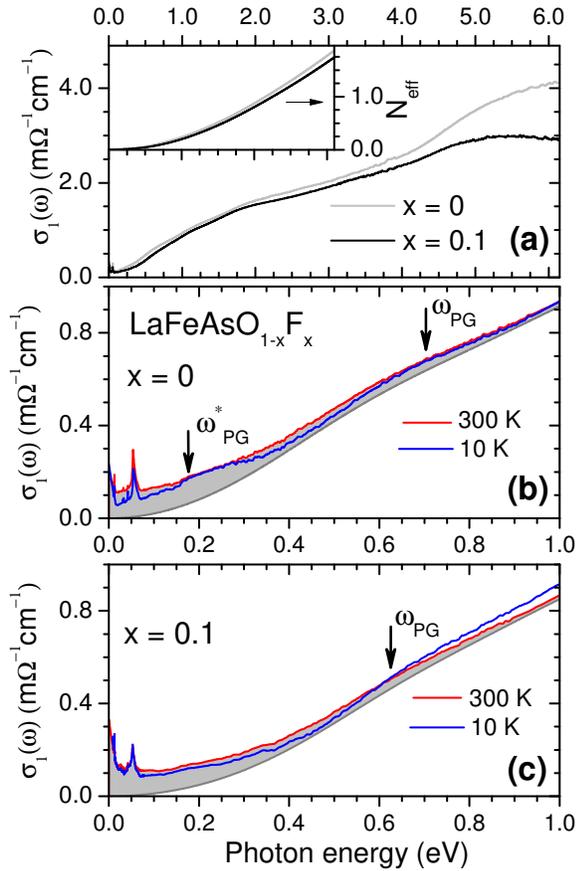}
\caption{(a) Optical conductivity of the LaFeAsO$_{0.9}$F$_{0.1}$ (black) and LaFeAsO (gray) measured at $T=300$. Low-energy part of
the spectra for (b) undoped and (c) F-doped samples at 300 K (red) and 10 K (blue). The dark gray lines and shaded areas in (b) and (c) show the low-energy
tail from the interband transitions and the remaining intraband charge carrier
contribution, respectively. Arrows mark the pseoudogap opening at $\omega_{PG}
\approx 0.65$ eV ($\omega^*_{PG}\approx 0.16$ eV).
 }
\protect\label{fig2} 
\end{figure}

The ellipsometric measurements in the frequency range 0.75 - 6.5 eV were performed with a  Woolam VASE variable angle ellipsometer of rotating-analyzer type. The sample was mounted on the cold finger of a helium flow cryostat with a base pressure of 2x10$^{-9}$ Torr at room temperature. For the IR measurements from 0.01 to 1.0 eV we used home-built ellipsometers in combination
with a Bruker IFS 66v/S FT-IR spectrometer. Some of the experiments were performed at the infrared beam line IR1 at the ANKA synchrotron in
Karlsruhe, Germany. The complex dielectric function, $\tilde \varepsilon(\omega) = \varepsilon_1(\omega)+\emph{i}\varepsilon_2(\omega)$, and the related optical conductivity $\sigma_1(\omega)=\omega \varepsilon_2(\omega)/(4\pi)$, were directly determined from the ellipsometric parameters $\Psi(\omega)$ and $\Delta(\omega)$ \cite{handbook}. The  inversion of the ellipsometric data was performed within the framework of an effective medium approximation (EMA), which in the case of polycrystallline samples corresponds to the volume average of the anisotropic dielectric tensor projections. We do not take into account possible perturbations of the ellipsometric data due to surface roughness or grain texturing. Our measurements at multiple angles of incidence from 55$^\circ$ to 85$^\circ$  confirm that the surface roughness effect on $\tilde \varepsilon(\omega)$ is less than 15$\%$ over the measured spectral range
and therefore does not significantly influence relative changes of $\tilde \varepsilon(\omega)$. Preferred orientation of the $ab$ planes parallel to the surface \cite{Hunte} could enhance the relative weight
of the in-plane components, but should not influence the overall distribution of spectral features. Moreover, as both samples were prepared in an identical fashion, their spectra can be directly compared irrespective of these possible distortions.

\begin{figure*}
\includegraphics[width=16cm]{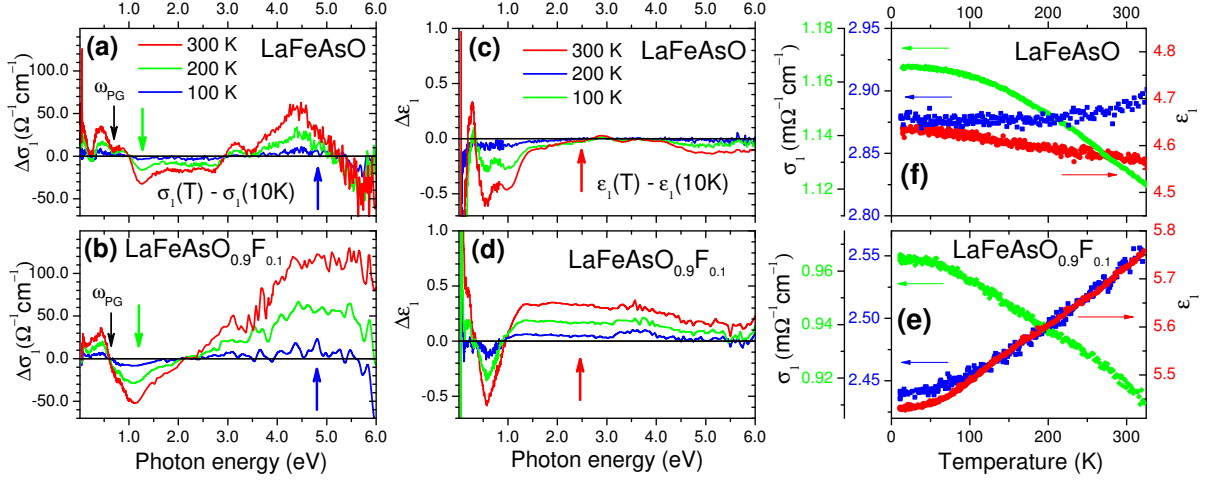}
\caption{Difference spectra (a,b) $\Delta\sigma_1(T,\omega)=\sigma_1(T,\omega)-\sigma_1(10K,\omega)$
and (c,d) $\Delta\varepsilon_1(T,\omega)=\varepsilon_1(T,\omega)-\varepsilon_1(10K,\omega)$
for LaFeAsO (upper panels) and LaFeAsO$_{0.9}$F$_{0.1}$ (lower panels). Black arrows in (a) and (b) indicate the same energy $\omega_{PG}\approx 0.65$ eV as in Figs. 2b,c.  (e,f) Temperature dependence of $\sigma_1(\omega)$ and $\varepsilon_1(\omega)$ measured at representative energies 1.2, 2.5 and 4.8 eV as marked by green, blue and red arrows in (a-d), respectively. Cooling-down and warming-up curves are consistent and averaged.   
 }
\protect\label{fig3} 
\end{figure*} 

Figure 1 shows the complex dielectric function of the LaFeAsO$_{0.9}$F$_{0.1}$ sample determined from the spectroscopic ellipsometry measurements at $T=300$ K. The optical response is due to strongly superimposed optical bands, which fall into low- and high-energies groups around $\sim$ 1 eV and $\sim$ 5 eV. In addition, there is a broad featureless contribution at intermediate
energies around $\sim$  3 eV. To separate contributions from the different bands, we performed a classical dispersion analysis. Using a dielectric function of the form $\tilde\varepsilon(\omega) =\varepsilon_{\infty}+ \sum_j
\frac{S_j}{\omega_j2-\omega2-i\omega\Gamma_j}$, where $\omega_j$,
$\Gamma_j$, and $S_j$ are the peak energy, width, and dimensionless
oscillator strength of the $j$th oscillator, and $\varepsilon_{\infty}$ is the core contribution from the dielectric function, we fit a set of Lorentzian oscillators simultaneously to $\sigma_1(\omega)$ and $\varepsilon_1(\omega)$. We introduced a minimum set of oscillators, with one oscillator beyond the spectral range investigated. Figure 1 summarizes results of this dispersion
analysis. One can clearly distinguish three low-energy optical bands
$\alpha_1$, $\alpha_2$, and $\alpha_3$, located at 0.75, 1.15, and
1.9 eV, and two high-energy optical bands $\gamma_1$ and $\gamma_2$,
located at 4.8 and 5.8 eV. The contribution at intermediate energies
is described by a single broad Lorentzian band $\beta$ peaked at 3.2
eV.  Based on a comparison of our data to band structure
calculations \cite{kotliar1,singh,dolgov}, the low-energy optical bands $\alpha_1$,
$\alpha_2$, and $\alpha_3$ can be assigned to a variety of
Fe$3d$-Fe$3d$ transitions, the broad contribution at intermediate
energies around $\beta$ to As$4p$-$Fe$3d interband transitions, and
the high-energy optical bands $\gamma_1$ and $\gamma_2$ to
O2$p-$Fe$3d$ and O$2p-$La$5d$ transitions, respectively.

Figure 2a shows the real part of the optical conductivity $\sigma_1(\omega)$ of both  samples in a wide spectral range up to 6 eV at $T=300$ K. F-substitution results in a noticeable decrease of the optical conductivity at energies above 4 eV, which may be related to the reduced O$2p$ character in this spectral range. However, F-doping does not strongly affect the response of
electronic charge carriers  near the Fermi level (Figs. 2b,c). The
far-IR optical conductivity level is at about 110 - 130 $\Omega^{-1}$cm$^{-1}$ for both samples. An upturn in the optical conductivity is only observed at very low energies $\leq 25$ meV, and in the zero-energy limit  the $\sigma_1(\omega)$ approaches the $dc$ conductivity of about 170 and 350 $\Omega^{-1}$cm$^{-1}$,
measured for the pristine and F-substituted samples, respectively. The inset in Fig. 2a shows the effective number of charge carriers per unit cell, $N_{eff}(\omega)$, participating in the optical excitations up to $\omega$ for both samples. This is estimated by integrating the optical conductivity using the relation
$N_{eff}(\omega)=\frac{2mV}{\pi e2}$SW$(\omega)$, where the spectral weight SW$(\omega)=\int_0^\omega \sigma_1(\omega^{'}) d\omega^{'}$, $m$ is the free electron mass and $V$ represents the unit-cell volume. By integrating over the spectral range where the dominant contribution comes from the three low-energy $\alpha$-optical bands, we determine $N_{eff}$(2.5 - 3 eV) $\sim$ 1.5 - 1.8 electrons per unit cell. We estimate the contribution from the free
electron charge carriers by subtracting contributions from phonons
and from the  interband  optical transitions (gray curves in
Figs. 2b,c) identified by our dispersion analysis. The gray-shaded area in Fig. 2b (Fig. 2c) gives the effective charge carrier density in the pure (F-substituted) sample $N_{eff}^D=0.043 (0.038)$, which is consistent with the plasma frequency $\omega_{pl}$ = 0.67 eV (0.61 eV) of  the  heavily damped
Drude term, $\gamma^D$ = 0.55 eV (0.47 eV), shown by the dark gray-shaded area in Fig. 1a.

The effective carrier density estimated from the optical response is
far below the corresponding values predicted by band structure
calculations, which show a large Fe$3d$ density of states at the
Fermi level \cite{singh,dolgov}. For example, considering the isotropic effective medium in a linear approximation, we estimate $N_{eff}^{EMA}$ = $\frac{mV}{4\pi e2}\frac{2\omega_{ab}2 + \omega_{c}2}{3}$ $\approx$ 0.37, using the anisotropic in-plane and out-of-plane plasma frequencies,
$\omega_{ab}$ = 2.30 eV and $\omega_{c}$ = 0.32 eV, determined from
the LDA calculations \cite{dolgov}. 
This may imply that the SW of the free charge carriers is strongly renormalized due to electron-lattice or/and electron-electron interactions. Assuming
that the electron-phonon interactions in LaFeAsO$_{1-x}$F$_x$ are
sufficiently strong, the SW will be shifted from the Fermi level to
polaronic excitations, superimposed with multiorbital Fe$3d$
transitions in the spectral range of $\alpha$- optical bands. In the
case of the strong electron-electron correlations in
LaFeAsO$_{1-x}$F$_x$, the conduction band will be strongly
renormalized while most of its SW is transferred into a
broad Hubbard band at higher energy $\sim U$ \cite{kotliar1,kotliar2}.

\begin{figure}[ht]
\includegraphics[width=7cm]{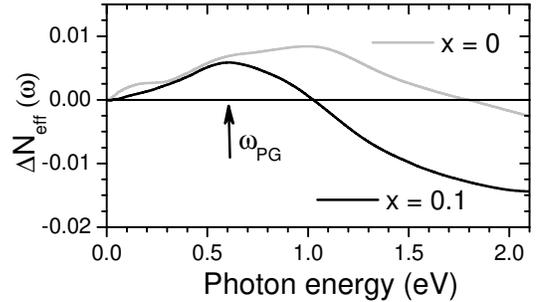}
\caption{Spectral weight changes $\Delta N_{eff}(\omega)=N_{eff}(\omega,300$
K$)-N_{eff}(\omega,10$ K$) $ in LaFeAsO$_{0.9}$F$_{0.1}$ (black) and LaFeAsO (gray). The black arrow marks the same energy $\omega_{PG}\approx 0.65$ eV as in Figs.2 b,c and Figs.3 a,b.}
\protect\label{fig4} 
\end{figure}

The temperature dependence of the optical SW can be instrumental in tracking down the origin of the crossover from the coherent Fermi liquid state to the incoherent regime in LaFeAsO$_{1-x}$F$_x$. The low-energy data of Figs.
2b,c already indicate the opening of the pseudogap with decreasing temperature at $\hbar \omega$ $\lesssim 0.65$ eV in both samples. In addition, the undoped LaFeAsO sample reveals gapping at lower energies  $\hbar
\omega$ $\lesssim 0.16$ eV and shows anomalies in the far-IR conductivity at the magnetic transition temperature $T_{SDW} \approx
155$ K. The details of the far-IR anomalies will be reported in a
forthcoming publication. In this Letter, we focus on the large-scale
SW transfer associated with the opening of the pseudogap. Figures 3a,b and c,d show temperature-difference spectra, $\Delta \sigma_1(\omega,T)$ and $\Delta \varepsilon_1(\omega,T)$ for the LaFeAsO and LaFeAsO$_{0.9}$F$_{0.1}$ samples, respectively, while Figs. 3e,f detail the temperature dependencies of $\sigma_1$ and $\varepsilon_1$ at representative photon energies.
Positive $\Delta \sigma_1(\omega)$ changes below 0.65 eV indicate
the evolution of the pseudogap with temperature. One can notice a
concomitant transfer of SW to the higher-energy range around 1.2 eV
(green arrows in Figs. 3a,b), coincident with the peak of the intense
$\alpha_2$ optical band in Fig. 1a. The green lines in Figs. 3e,f show
that these changes are continuous, with no singularity at any
temperature in both samples.

Figure 4 shows the integrated changes in $\sigma(\omega)$ between 300 and 10 K in terms of $\Delta N_{eff}(\omega)$. Apparently, the SW loss, associated with the opening of the pseudogap in LaFeAsO is well balanced by the SW accumulated in a narrow energy range above the pseudogap, resulting in $\Delta N_{eff}(2.0 eV)\approx 0$ (gray line in Fig. 4). The additional features in $\Delta \sigma_1(\omega,T)$ above this energy arise from the conventional temperature behavior of the interband optical transitions ($\beta$ and $\gamma$), that is, an intraband redistribution of the SW due to narrowing and blue-shifts of the individual optical bands upon cooling. The $T$-dependence of the $\varepsilon_1(\omega,T)$ affords an independent and complementary probe of the SW shift. Figure 3c
and the red line in Fig. 3f show that $\varepsilon_1(\omega,T)$ is nearly temperature independent between 2.0 and 4.0 eV, which confirms the lack of any SW transfer between the $\beta$- and $\gamma$- optical bands and lower energies. In clear contrast, $\varepsilon_1(\omega,T)$ of the LaFeAsO$_{0.9}$F$_{0.1}$ sample exhibits strong temperature dependence at these energies, as shown
by Figs. 3d,e. This strong positive shift of $\varepsilon_1$ with
temperature is indicative of a substantial SW transfer from low
($\hbar \omega \lesssim 2.0$ eV) to high ($\hbar \omega \gtrsim 4.0$
eV) energies. The blue line in Fig. 3e shows the corresponding
increase in $\sigma_1$ at 4.8 eV with temperature. Thus, in LaFeAsO$_{0.9}$F$_{0.1}$ the SW gain at $\hbar \omega_{PG} \lesssim$ $\hbar \omega \lesssim$ 2.0 eV exceeds the SW loss below $\omega_{PG}$ by $\Delta N_{eff} \approx$ 0.015, as shown in Fig. 4. This extra SW comes from higher energies and accounts for 40 \% of the charge carriers SW $N^D_{eff} \approx$ 0.038 electrons per cell or 1.5 \% of the total SW below 2.0 eV (inset of Fig. 2a). 

A similar variation as large as $4-5\%$ of the total SW integrated below 2.0 eV has been observed in the normal state of high-$T_c$ cuprates and quantitatively described by DMFT calculations, with the introduction of strong correlation effects \cite{kotliar3,toschi}. By this approach, the temperature dependence of the SW is controlled by renormalized quasiparticle dispersion near the Fermi level. The temperature-driven SW transfer observed in the present study shows no indication of a discontinuity near $T_c$, and thus we cannot unequivocally accredit the observed effect with a normal state of the superconducting phase of LaFeAsO$_{0.9}$F$_{0.1}$. Nevertheless, this observation is a definitive signature of electronic correlations in Fe-based oxypnictides, which means that the optical conductivity above $\sim$ 4 eV includes optical transitions from the lower Hubbard band to the quasiparticle peak above the Fermi level \cite{kotliar1}.  To verify if the anomalous F-doping effect on the temperature dependence of the SW is inherent to the superconducting phase, further  measurements on single crystals are needed. This would further imply, if confirmed, that doping leads to a dramatic change in the electron dispersion at the Fermi level. Actually, that  is very likely in multiorbital LaFeAsO, where the Fermi surface topology is expected to be easily modified by doping \cite{mazin,yin}.

To summarize, we report temperature dependences of the optical dielectric response of the superconducting LaFeAsO$_{0.9}$F$_{0.1}$ ($T_c$ = 27 K) and nonsuperconducting LaFeAsO polycrystalline samples  over a wide range of temperatures (10 - 350 K) and photon energies (0.01 - 6.5 eV). Our optical
conductivity spectra provide evidence that LaFeAsO$_{1-x}$F$_{x}$ is a low-carrier
density metal and may imply that the electronic states near the Fermi surface are strongly renormalized due to electron-phonon and/or electron-electron interactions. Further support for strong electronic correlations is derived
from the temperature and doping dependence of the high-energy optical response.

\begin{acknowledgments}
We thank L. Boeri, O.V. Dolgov, I.I. Mazin, A. Toschi, A.N. Yaresko, O.K.
Andersen for helpful discussion, Y.-L. Mathis for his support at ANKA.
\end{acknowledgments}

\bibliographystyle{apsrev} 
\bibliography{lofa_vis_ab}
\end{document}